\documentclass[api,rsi,floatfix,superscriptaddress,reprint]{revtex4-1}
\usepackage{graphicx}
\usepackage{amsmath}
\usepackage{amsfonts}
\usepackage{amsthm}
\usepackage{float}
\usepackage[margin=1in]{geometry}
\usepackage{enumerate}
\usepackage{etoolbox}
\usepackage{multirow}
\usepackage{amssymb}
\usepackage{comment}
\usepackage{mathrsfs}
\usepackage{ gensymb }
\usepackage{braket}
\usepackage[font=small, format=plain,justification=centerlast]{caption}

\begin{document}
\title{Decoupling of a Neutron Interferometer from Temperature Gradients}

\author{P. Saggu}
\affiliation{Department of Chemistry, University of Waterloo, Waterloo, ON, Canada, N2L3G1}
 
\author{T. Mineeva} 
\email[Corresponding author: ]{taisiya.mineeva@uwaterloo.ca}
\affiliation{Department of Physics and Astronomy, University of Waterloo, Waterloo, ON, Canada, N2L3G1}
\affiliation{Institute for Quantum Computing, University of Waterloo,  Waterloo, ON, Canada, N2L3G1}

\author{M. Arif}
\affiliation{National Institute of Standards and Technology, Gaithersburg, MD 20899, USA}

\author{D. G. Cory}
\affiliation{Department of Chemistry, University of Waterloo, Waterloo, ON, Canada, N2L3G1}
\affiliation{Institute for Quantum Computing, University of Waterloo,  Waterloo, ON, Canada, N2L3G1} 
\affiliation{Canadian Institute for Advanced Research, Toronto, Ontario M5G 1Z8, Canada}
\affiliation{Perimeter Institute for Theoretical Physics, Waterloo, ON, Canada, N2L2Y5}

\author{R. Haun}
\affiliation{Department of Physics, Tulane University, New Orleans, LA 70118, USA}
\author{B. Heacock}
\affiliation{Department of Physics, North Carolina State University, Raleigh, NC 27695, USA}
\affiliation{Triangle Universities Nuclear Laboratory, Durham, North Carolina 27708, USA}

\author{M. G. Huber}
\email[Corresponding author: ]{michael.huber@nist.gov}
\affiliation{National Institute of Standards and Technology, Gaithersburg, MD 20899, USA}

\author{K. Li}
\affiliation{Department of Physics, Indiana University, Bloomington, Indiana 47408, USA}
\affiliation{Center for Exploration of Energy and Matter, Indiana University, Bloomington, IN 47408, USA}

\author{J. Nsofini} 
\affiliation{Department of Physics and Astronomy, University of Waterloo, Waterloo, ON, Canada, N2L3G1}
\affiliation{Institute for Quantum Computing, University of Waterloo,  Waterloo, ON, Canada, N2L3G1}

\author{D. Sarenac}
\affiliation{Department of Physics and Astronomy, University of Waterloo, Waterloo, ON, Canada, N2L3G1}
\affiliation{Institute for Quantum Computing, University of Waterloo,  Waterloo, ON, Canada, N2L3G1} 
\author{C. B. Shahi} 
\affiliation{Department of Physics, Tulane University, New Orleans, LA 70118, USA}

\author{V. Skavysh}
\affiliation{Department of Physics, North Carolina State University, Raleigh, NC 27695, USA}

\author{W. M. Snow}
\affiliation{Department of Physics, Indiana University, Bloomington, Indiana 47408, USA}
\affiliation{Center for Exploration of Energy and Matter, Indiana University, Bloomington, IN 47408, USA}
\author{S. A. Werner}
\affiliation{Physical Measurement Laboratory, National Institute of Standards and Technology, Gaithersburg, MD 20899, USA}

\author{A. R. Young}
\affiliation{Department of Physics, North Carolina State University, Raleigh, NC 27695, USA}
\affiliation{Triangle Universities Nuclear Laboratory, Durham, North Carolina 27708, USA}

\author{D. A. Pushin}
\email[Corresponding author: ]{dmitry.pushin@uwaterloo.ca}
\affiliation{Department of Physics and Astronomy, University of Waterloo, Waterloo, ON, Canada, N2L3G1}
\affiliation{Institute for Quantum Computing, University of Waterloo,  Waterloo, ON, Canada, N2L3G1}

\begin{abstract}
Neutron interferometry enables precision measurements that are typically operated within elaborate, multi-layered facilities which provide substantial shielding from environmental noise. These facilities are necessary to maintain the coherence requirements in a perfect crystal neutron interferometer which is extremely sensitive to local environmental conditions such as temperature gradients across the interferometer, external vibrations, and acoustic waves. The ease of operation and breadth of applications of perfect crystal neutron interferometry would greatly benefit from a mode of operation which relaxes these stringent isolation requirements. Here, the INDEX Collaboration and National Institute of Standards and Technology demonstrates the functionality of a neutron interferometer in vacuum and characterize the use of a compact vacuum chamber enclosure as a means to isolate the interferometer from spatial temperature gradients and time-dependent temperature fluctuations. The vacuum chamber is found to have no depreciable effect on the performance of the interferometer (contrast) while improving system stability, thereby showing that it is feasible to replace large temperature isolation and control systems with a compact vacuum enclosure for perfect crystal neutron interferometry.
\end{abstract}

\maketitle
\section{Introduction}
Since the first demonstration of a perfect crystal neutron interferometer (NI) in 1974 \cite{RAUCH}, numerous experiments have been performed exploring both the nature of the neutron and its interactions \cite{Shull1968,Sam}. Examples include the first observation of a quantum phase shift due to the Earth's gravitational field \cite{COW}, demonstrating the $4\pi$ symmetry of spinor rotation \cite{spin}, and measuring the effect of the Earth's rotation on the phase of the neutron (Sagnac effect) \cite{SAGNAC} among other fundamental experiments. In addition, neutron interferometry has been used for precision measurements of coherent and incoherent neutron-nucleus scattering cross sections \cite{Schoen,Huffman, Huber}. 


 
\begin{figure}
\centering
\includegraphics[width=0.98\columnwidth]{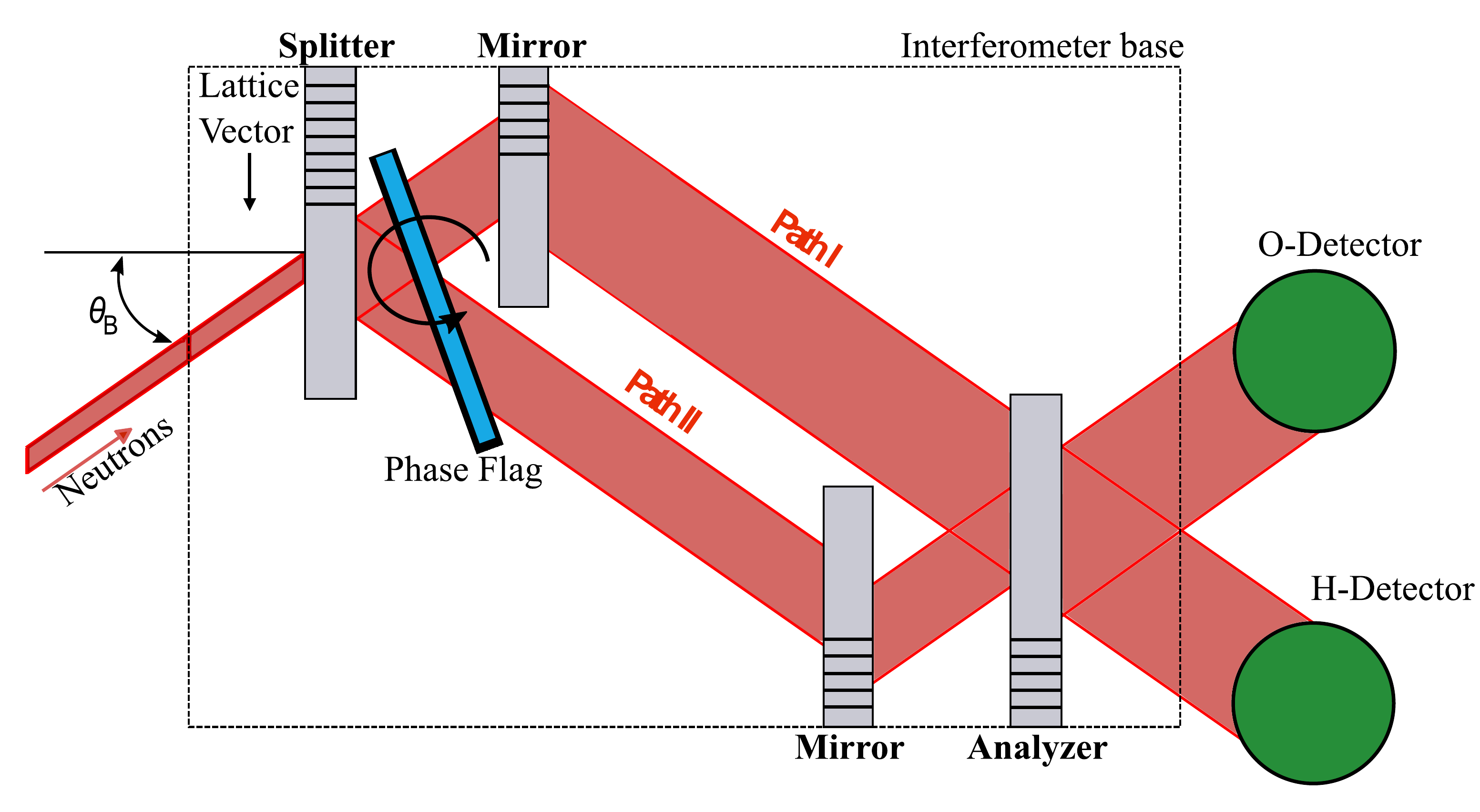}
\caption{ Schematic diagram of a perfect crystal silicon skew-symmetric neutron interferometer. This interferometer provides two widely separated coherent beams, labeled Path I and Path II. A fused silica phase flag in the beam paths is rotated to produce changes in the neutron path lengths. 
}
\label{fig:NI}
\end{figure}

A NI is cut from a float-zone grown silicon ingot so that a series of Bragg-diffracting ``blades" protrude from a common base (Figure~\ref{fig:NI}).  The first blade of the NI creates a coherent superposition of two spatially separated paths; Path I and Path II. Intermediate blades act as mirrors, and the final, analyzing blade coherently mixes Paths I and II, forming an interference pattern in two exit beams labeled the O- and H-beams. The arrangement is analogous to a Mach-Zehnder optical interferometer ~\cite{PushinAHEP}. 
 

Upon exiting the NI, the neutrons are detected with nearly 100 \% efficiency by $^{3}$He gas filled proportional detectors. In order to observe interference patterns at the detectors, a phase difference between the two NI paths must be introduced and modulated. This is done by placing and rotating a material, with an index of refraction different than air (referred to as a \textit{phase flag}), between the interferometer blades. The material used for a phase flag is typically an optically flat sample of fused silica supported above the interferometer using a steel rod. Neutron intensities at the detectors measured as a function of phase flag rotation angle produce an interference pattern commonly referred to as an interferogram or a contrast curve. As the neutron absorption in silicon is negligible, the sum of the O- and H-beam intensities is a constant.

 
The intensity at the O-detector is given by: 
\begin{align}
I_{O}&= \textrm{A}+ \textrm{B cos}[g(\epsilon) + \phi_{0}]
\label{eqn:INT}
\end{align}
where $A$ and $B$ are fit parameters, g($\epsilon$) is the action of the phase flag at angle $\epsilon$, and $\phi_{0}$ is an intrinsic phase shift. The intrinsic phase shift is the experiment-dependent, relative phase difference between the two neutron paths and is typically a parameter of interest in a neutron interferometry experiment.

The contrast is a measure of the degree of coherence between the two paths inside the interferometer. Contrast (C) at the O-beam detector is determined from the ratio of the fit parameters (Equation~\ref{eqn:INT}) and reflects the depth of the intensity modulations: 
 \begin{align}
C&=\frac{\textrm{B}}{\textrm{A}}
\end{align}

While for an ideal interferometer the contrast is 100 \%, this is never achieved in practice for a number of reasons.  Some of these contrast loss mechanisms come from surface and crystallographic imperfections of the phase flag and the NI blades and are therefore associated with the interferometer itself and not the environment. Machining of interferometers is done with a diamond coated wheel, ensuring that the surfaces of the blades are parallel within a tolerance of a few $\mu$m. This is typically followed by chemical etching to remove any residual strain in the crystal lattice. The presence of impurity in the silicon crystal may introduce distortions that spoil the parallelism of the lattice planes~\cite{Sam}. In the case of a wide neutron beam that interacts with a larger section of the interferometer blade, contrast is reduced compared to the optimally located narrow focused beam.  Other contrast loss mechanisms originate from changes in the external environment such as acoustic waves, vibrations, temperature gradients, and the humidity of the air, all of which can couple to the NI crystal geometry and internal conditions. Many of these effects could be suppressed or eliminated if one operated the NI in vacuum.  

The extreme sensitivity of the NI to strain and crystal imperfections, a historical focus on damping vibrational noise, the negligible small phase shift due to air, and accessibility are major reasons why vacuum enclosures have never been explored before in neutron interferometry experiments. However, a temporal drift in temperature gradients within a NI changes the intrinsic phase shift.  Uneven thermal expansion of different parts of the interferometer leads directly to phase instabilities that limit the accuracy of precision measurements taken over long time periods. If these effects dominate, it should be possible to improve the interferometer performance by isolating the NI from air and from thermal contact from uncontrolled thermal surfaces, taking advantage of the high thermal conductivity of the silicon, and by establishing an isotropic, uniform thermal environment with thermal exchange dominated by radiation and by temperature-controlled surfaces. Here we demonstrate that enclosing an entire NI in vacuum is a space-efficient way to increase phase stability, without a sacrifice in contrast.

\maketitle
 \section{Neutron Interferometry and Optics Facilities at NIST}
 
The experiment described here was performed at the National Institute for Standards and Technology (NIST) Center for Neutron Research (NCNR). The neutron guide hall at the NCNR houses two Neutron Interferometry and Optics Facilities: NIOF~\cite{hutch} built in 1994 and  NIOFa~\citep{NIOFa} built in 2010. Each of these facilities utilizes separate beam lines delivering cold neutrons from a 20 MW reactor.

The original facility, NIOF, was built to preserve long-term phase stability. It is a massive enclosure that uses a \textquotedblleft box within a box within a box" approach \cite{ALLMAN}. The outermost enclosure is a concrete blockhouse that provides environmental isolation. It encompasses another massive enclosure that provides passive thermal and acoustic isolation. Finally, the innermost enclosure is a cadmium lined aluminum box that houses the interferometer and detectors. This enclosure is temperature controlled through a set of heaters and is able to maintain a constant internal temperature to within 5 mK~\citep{DIMAVCL}. Furthermore, the NIOF is constructed on its own foundation separate from the rest of the building and incorporates a state-of-the-art vibration isolation system \cite{ARIF,ALLMAN,RAKHECHA,BAUSPIESS}. With its ability to stabilize the temperature and damp vibrations, the original facility provides exceptional phase stability and high contrast for interferometry experiments. 


The NIOFa was built alongside the existing facility on cold guide NG7. It is a compact facility with increased accessibility that delivers a higher neutron flux. Its construction was in part motivated by recent advances in quantum information processing. The novel design of a decoherence free subspace (DFS) NI was shown to be robust against vibrational frequencies up to 22 Hz \citep{DIMA,DIMAHINDAWI}, therefore diminishing the necessity for traditional massive vibrational isolation enclosures.  

The environmental shielding at NIOFa largely differs from the design of the original facility. Its main wall consists of tall paraffin wax and steel shot-filled-steel walls surrounding the neutron guide. The interferometer's only enclosure is a cubic aluminum box with a length of 76.2 cm per side. The box is similarly lined with cadmium to decrease neutron background and sits on a 3 mm thick fiberglass base that isolates it thermally from the optical table. The compactness of the NIOFa facility comes at the expense of long-term phase stability and requires novel methods of environmental isolation. In this paper, we describe thermal isolation  of the neutron interferometer by using a vacuum chamber (Figure~\ref{fig:chamber}) which shields the interferometer from temperature fluctuations inside the guide hall.

\section{Vacuum Chamber}
\subsection{Design}
The vacuum chamber was constructed from Aluminum 6061-T6 because of its low neutron scattering and absorption cross-sections as well as its excellent thermal conductivity (167~W/(m$\cdot$K)). The body of the vacuum chamber is an aluminum tube with an inner diameter of 30.5~cm, an outer diameter of 32.5~cm, and an inner height of 20.3~cm giving a total enclosed volume of roughly 14830~cm$^3$. The top and bottom plates along with flanges were designed with grooves appropriate for Fluoropolymer O-rings. Flanges were welded onto the aluminum tube and were then screwed into upper and lower aluminum plates to close the chamber.  

\begin{figure}
\center
\includegraphics[width=0.98\columnwidth]{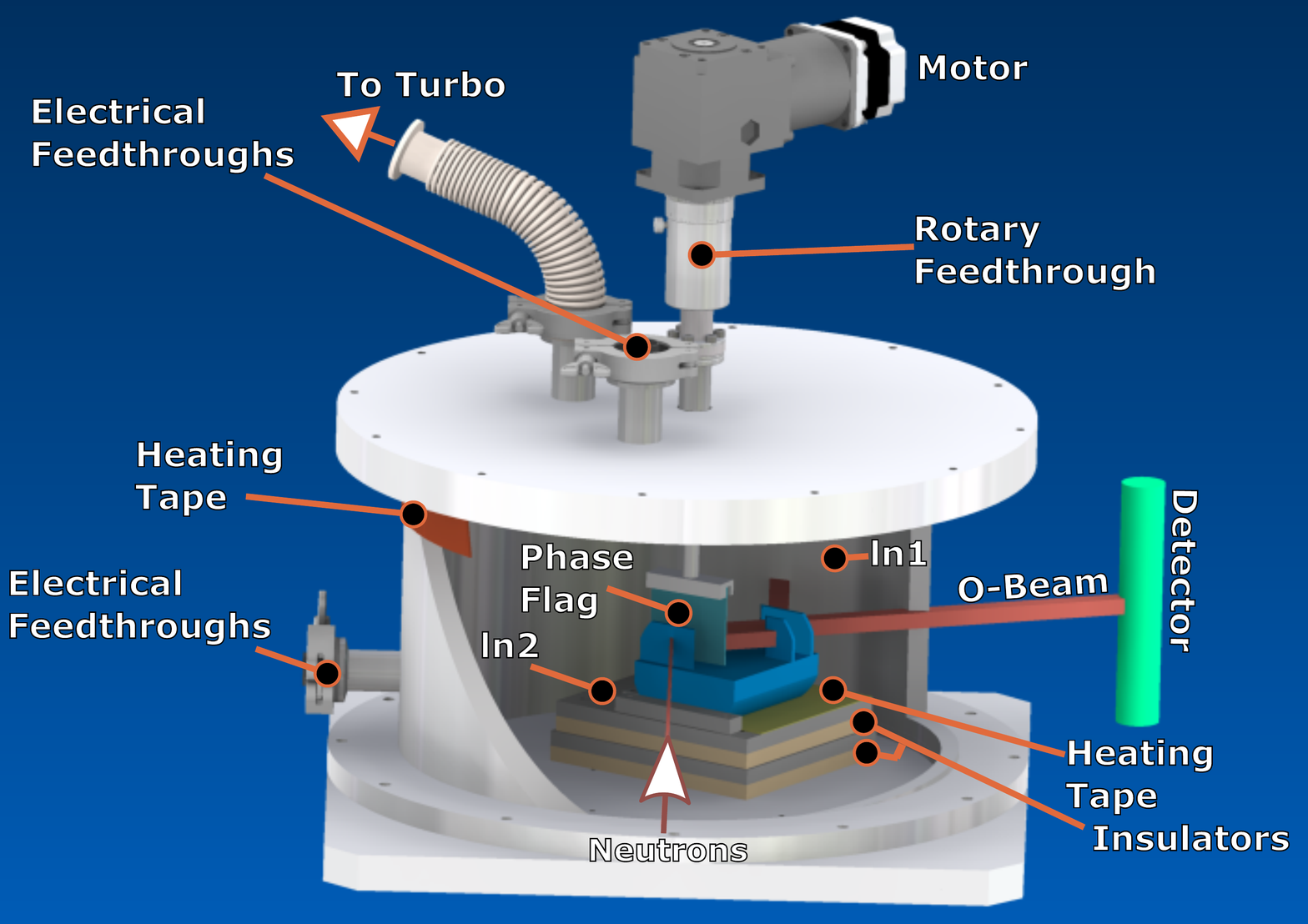}
\caption{Cross-sectional sketch of the vacuum chamber with the interferometer inside }
\label{fig:chamber}
\end{figure}

The vacuum chamber (Figure~\ref{fig:chamber}) was placed inside of the cubic, cadmium-lined aluminum box. The chamber itself was supported by the rotation, translation and height stages for optical alignment of the NI with the neutron beam. The interferometer was placed inside the chamber on a specially machined aluminum cradle. A polyoxymethylene plate was sandwiched between the base of the cradle and the vacuum chamber, secured using screws made from an organic thermoplastic polymer of polyether ether ketone (PEEK). Both Polyoxymethylene and PEEK were selected as insulators to prevent heat transfer between the vacuum chamber and the interferometer because of their good mechanical properties. To minimize temperature gradients inside the chamber two heating elements were installed: an internal one located below the interferometer base (denoted as "In2" on Figure 2) and an external one that was wrapped around the outside of the vacuum chamber walls (denoted as "In1" Figure~\ref{fig:chamber}). 


There are a number of vacuum compatible feedthroughs on the vacuum chamber. The first one is a rotary feedthrough to hold and rotate the phase flag. The second provides a connection to a vacuum pump while the third and fourth provide electrical connections for thermistors taped to the inner wall of the vacuum chamber and to the heater on the interferometer cradle. All electrical leads are twisted pairs to minimize stray magnetic fields. The two probes inside the vacuum chamber monitor the temperature difference between the wall of the chamber and the interferometer.  A third probe, a Pt~100 resistor, was taped to the outside of the vacuum chamber to gauge the overall temperature insulation provided by the chamber (Figure~\ref{fig:chamber}).  

\subsection{Numerical Simulations}
3D numerical simulations of the setup were done using finite element analysis  to gain insight into the temperature variation across the interferometer. The simulation geometry closely resembles the setup shown in Figure~\ref{fig:chamber}, and consists of the vacuum chamber, heating elements, skew symmetric NI, and the depicted cradle under the NI. Actual dimensions and material properties of those elements were used. Since the NI sits in the middle of vacuum chamber with two heating elements, one around the outside of the chamber walls and one directly under the interferometer base, both radiating and conducting heat transfer mechanisms were considered in our simulations. Initial simulations show that in order to maintain a stable temperature at the interferometer, it needs to be in contact with the heating element via a good thermal conductor. For that reason, we removed the insulating pool felt that has been historically placed under the NI base. 

\begin{figure}[ht]
\includegraphics[width=0.98\columnwidth]{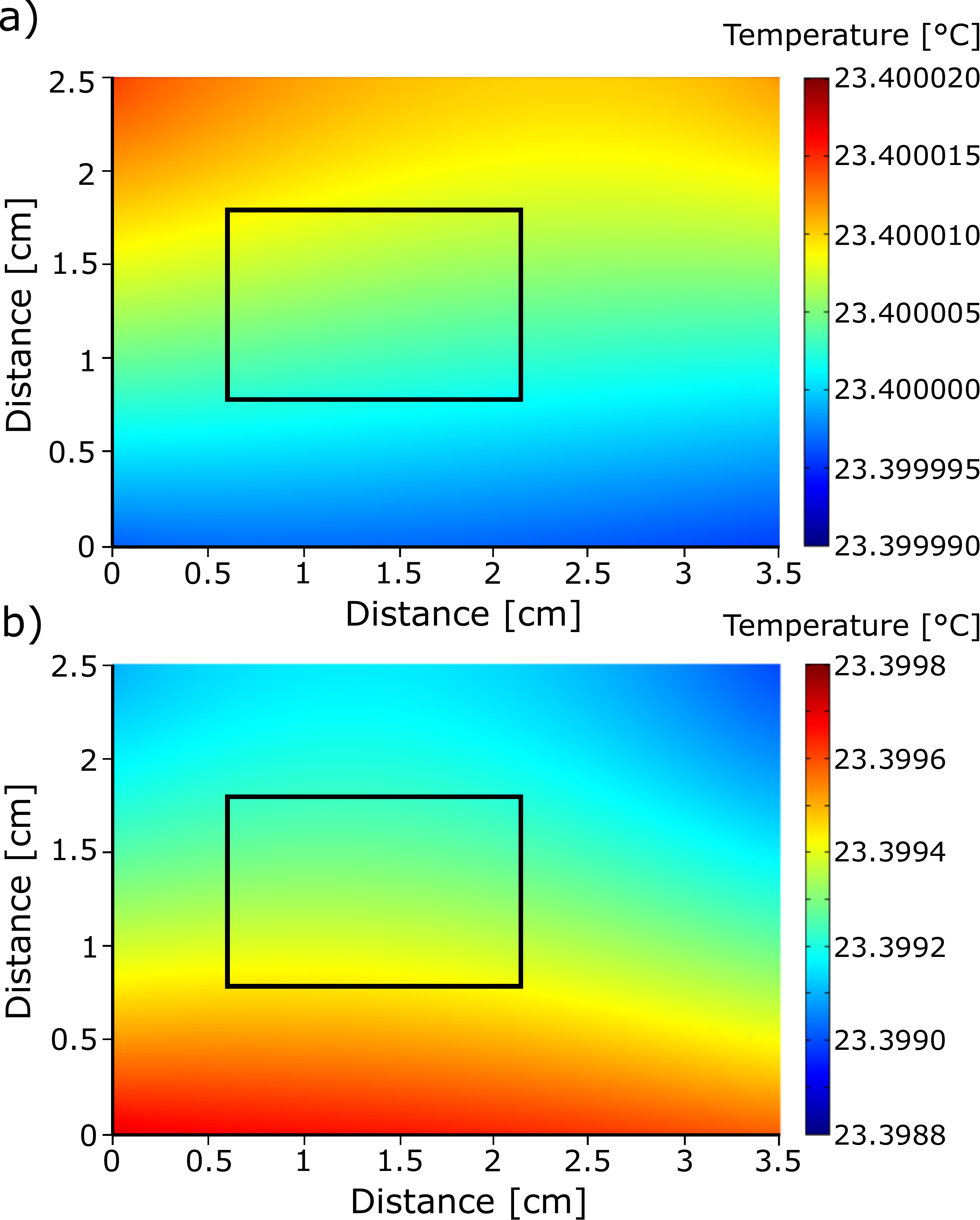}
\caption{Simulated temperature across the last NI blade given the setup in Figure~\ref{fig:chamber} with the heating element on the wall of the vacuum chamber set to $23.5~^\circ$C and the heating element at the base of interferometer set to $23.4~^\circ$C. The black rectangle depicts the expected neutron beam size. (a) The temperature variance across the blade given an ambient temperature of $23~^\circ$C and (b) given an ambient temperature of $19~^\circ$C. The simulated temperature variances are 0.00003 $^\circ$C and 0.001 $^\circ$C respectively.}
 \label{fig:simulations}
\end{figure}

Figure~\ref{fig:simulations} depicts the simulated temperature within the central cross section of the analyzer blade when the heating elements for the wall of the vacuum chamber and the base of interferometer were set to fixed temperatures of 23.5 $^\circ$C and 23.4 $^\circ$C respectively. The choice of the set temperatures of the heating elements in the simulations reflects experimentally determined temperature combinations for which phase stability was achieved.  

Figure~\ref{fig:simulations} (a) shows that for an external ambient temperature of $23~^\circ$C in such a configuration, the temperature across the last blade of the NI varies by 30 $^\circ\mu$C; while Figure~\ref{fig:simulations} (b) shows that for an external ambient temperature of 19 $^\circ$C the temperature varies by 1 $^\circ$mC. This is a factor of 100 more variation caused by a shift of only 4 $^\circ$C external. The black rectangle specifies the expected neutron beam size at the last NI blade, while the external temperature range reflects the typical temperature variance in the guide hall throughout the year.  
 
\section{Results} 
\begin{figure}[ht]
\center
\includegraphics[width=0.98\columnwidth]{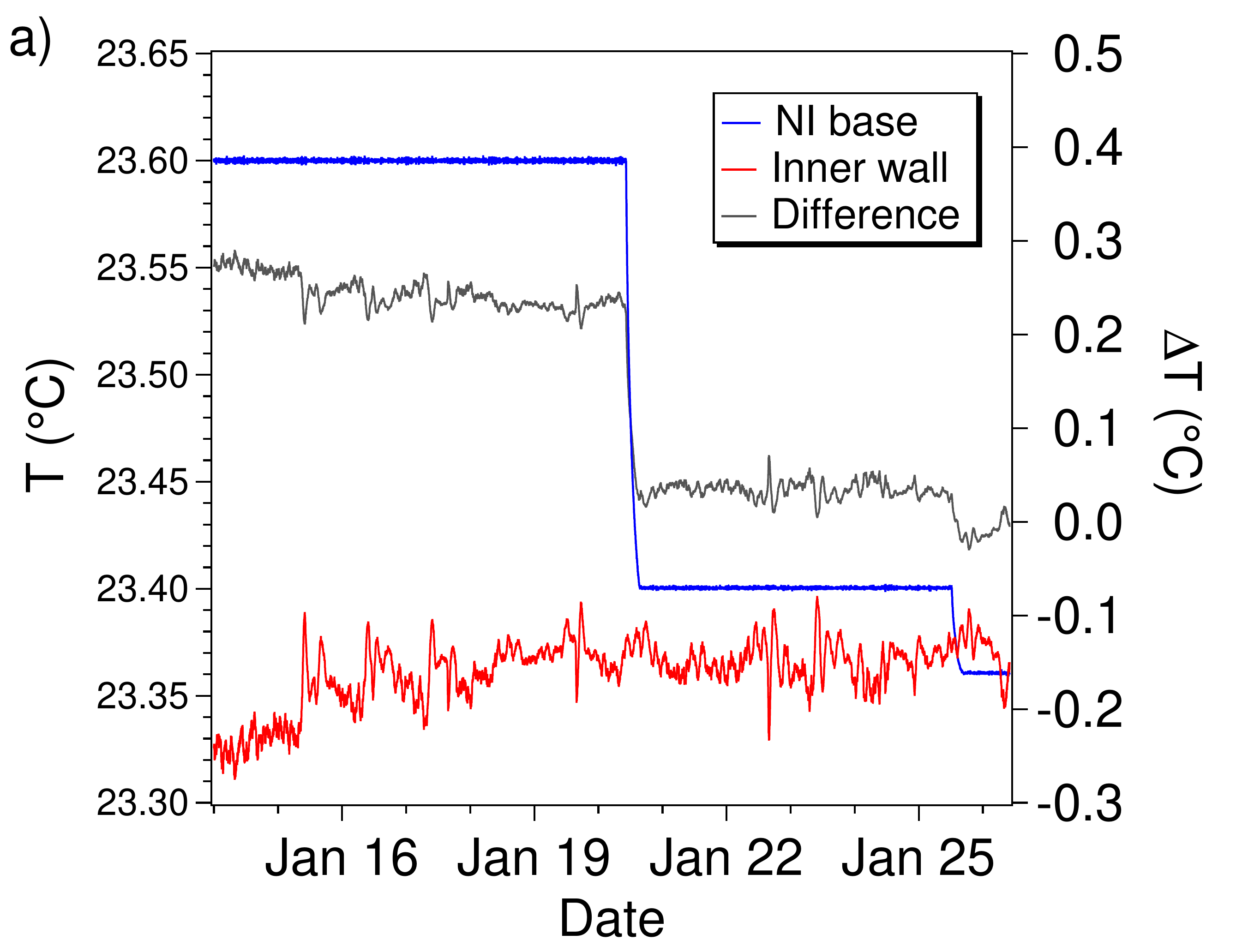}
\includegraphics[width=0.93\columnwidth]{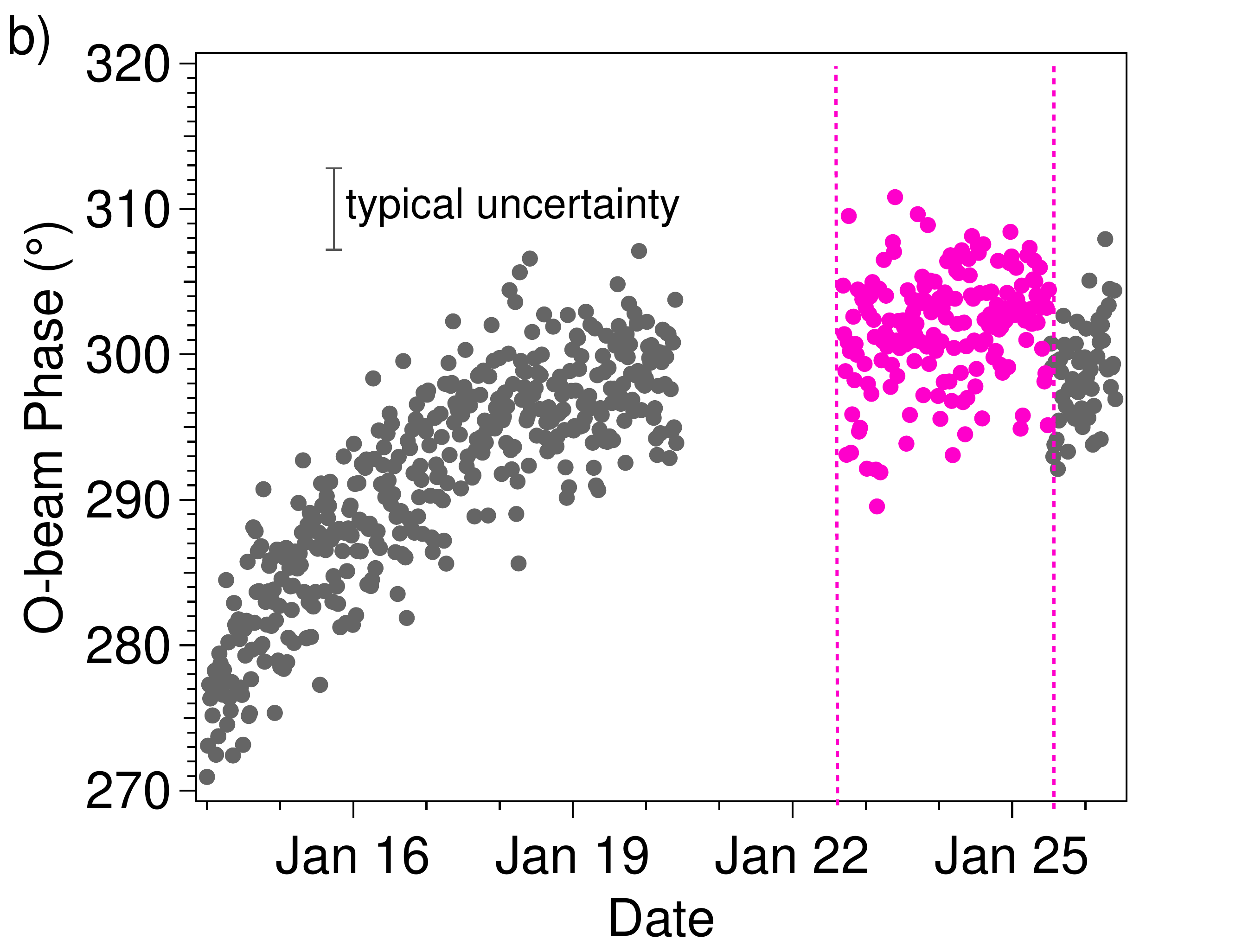}
\caption{(a) Temperature measurements at the inner chamber wall using ln1 (red, scale on left), interferometer base using ln2 (blue, scale on left) and their difference (grey, scale on right) (b) Corresponding phase measurements. The magenta region highlighted by the dotted lines is expanded in Figure~\ref{fig:Zoom}. The gap in the phase measurements between January 20$^\mathrm{th}$ and 22$^\mathrm{nd}$ stems from o reactor shut-down. A typical phase uncertainty is shown. This uncertainty is purely statistical and  at the 68 \% confidence level.}
 \label{fig:ItensitySPP}
\end{figure}
\begin{figure}[ht]
\center
\includegraphics[width=0.98\columnwidth]{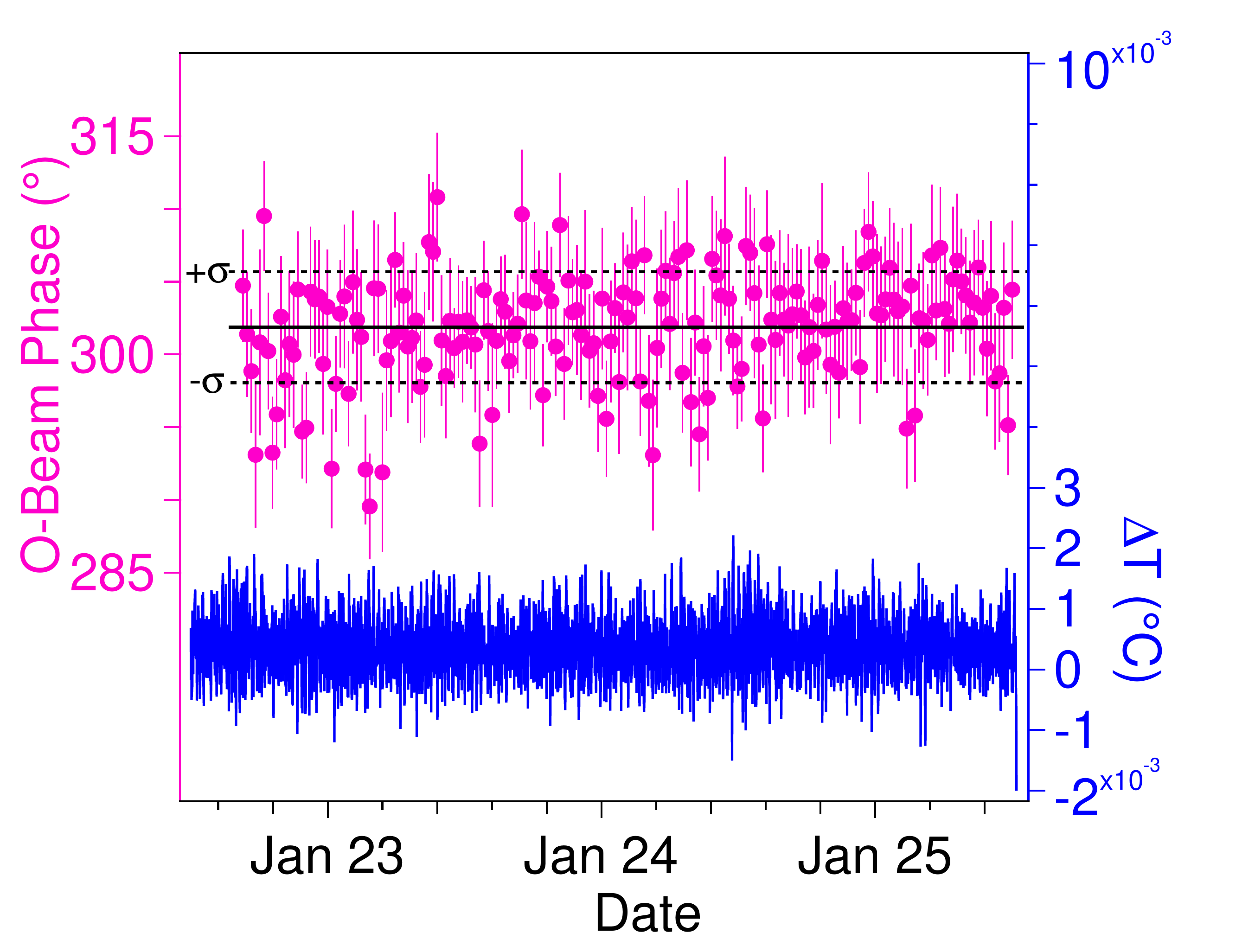}
\caption{Phase measurements over a period of 3 days (magenta) and corresponding temperature $\Delta T = (T-23.400)$ \degree C at the interferometer base (blue). Temperature stability improves NI phase stability.}
 \label{fig:Zoom}
\end{figure}
\begin{figure}[ht]
\center
\includegraphics[width=0.98\columnwidth]{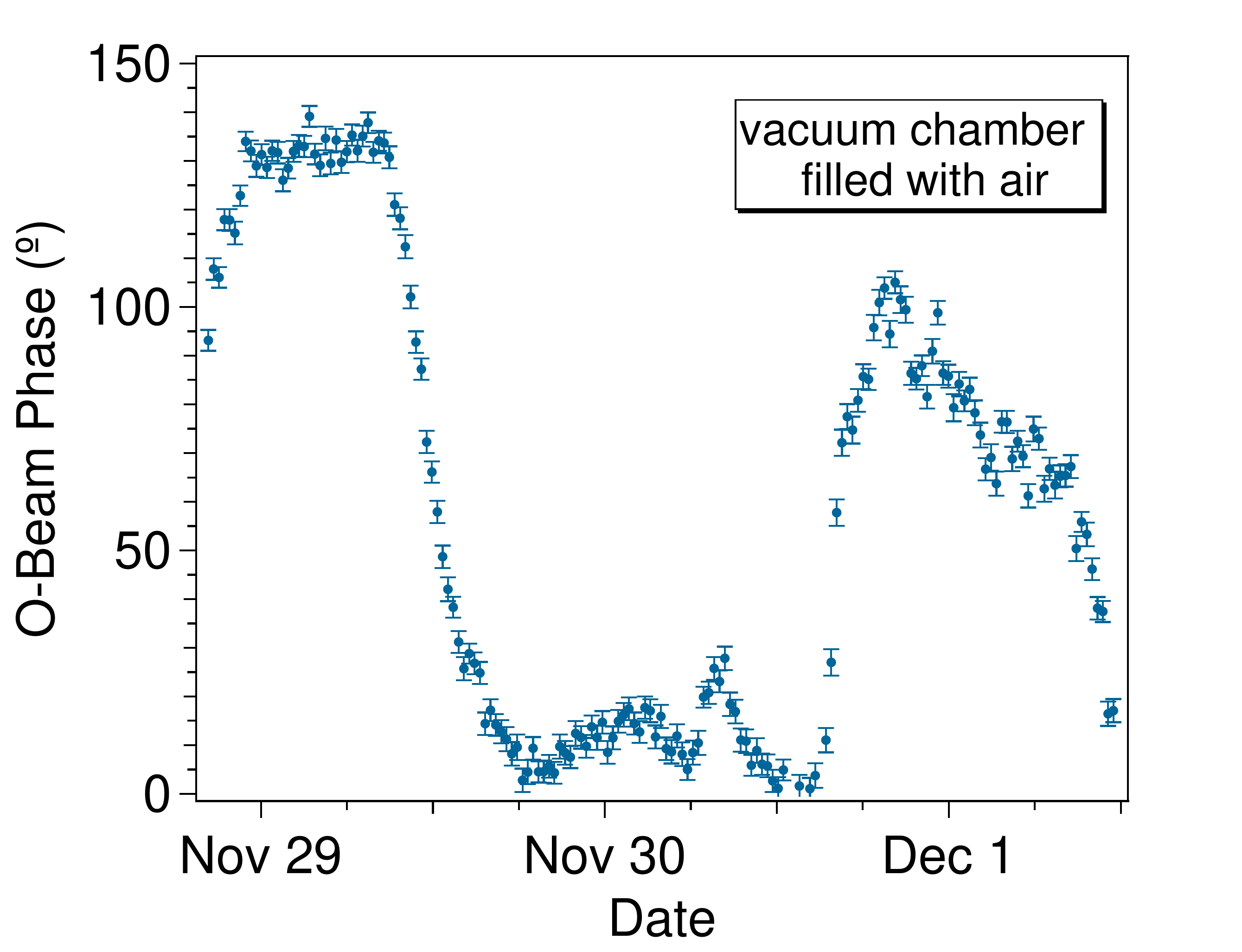}
\caption{ Phase measurements over a period of 3 days with no vacuum.  Phase uncertainties are purely statistical at the 68 \% confidence level.}
 \label{fig:phaseair}
\end{figure}

The vacuum system was tested at the new beam line (NIOFa) using a four blade skew-symmetric NI and a 2.2 \AA  ~incident neutron wavelength. Data for phase stability was collected over a period of several months. From Jan $15^{th}$ to Jan $20^{th}$ the heating element at the wall of the vacuum chamber was set to 23.5 $^\circ$C and the heating element at the base of interferometer was set to 23.6 $^\circ$C. The temperatures were measured using two thermistors, placed near the NI base and on the inner chamber wall. Figure~\ref{fig:ItensitySPP} shows the temperature and phase measurements for a subset of our data. The temperature difference between the base of the NI and the inner wall of the chamber was around 0.4 $^\circ$C and the phase drifted by approximately 30 $^\circ$ over a span of 5 days.  
  
We  achieved higher phase stability as we reduced the set point of the heating tape at the base of the NI to 23.4 $^\circ$C so that the temperature difference between the inner vacuum chamber walls and the NI base was minimized. This is shown in Figure~\ref{fig:Zoom} which expands on the temporal region indicated by dashed lines in Figure~\ref{fig:ItensitySPP}.  The temperature at the base of the interferometer was well stabilized between 23.400 \celsius~and 23.401 \celsius. During a 3 day period, we observed high phase stability, with no appreciable drift.  A linear fit of the data depicted in Figure~\ref{fig:Zoom} obtains a slope of only $(0.003 \pm 0.03)$ deg/hr at the 68 \% confidence level.
For reference, we collected phase data without the use of the vacuum chamber, and observed a peak-to-peak phase drift of 140$^\circ$ over a similar span of 3 days (Figure~\ref{fig:phaseair}).
  
\begin{figure}[ht]
\center
\includegraphics[width=0.98\columnwidth]{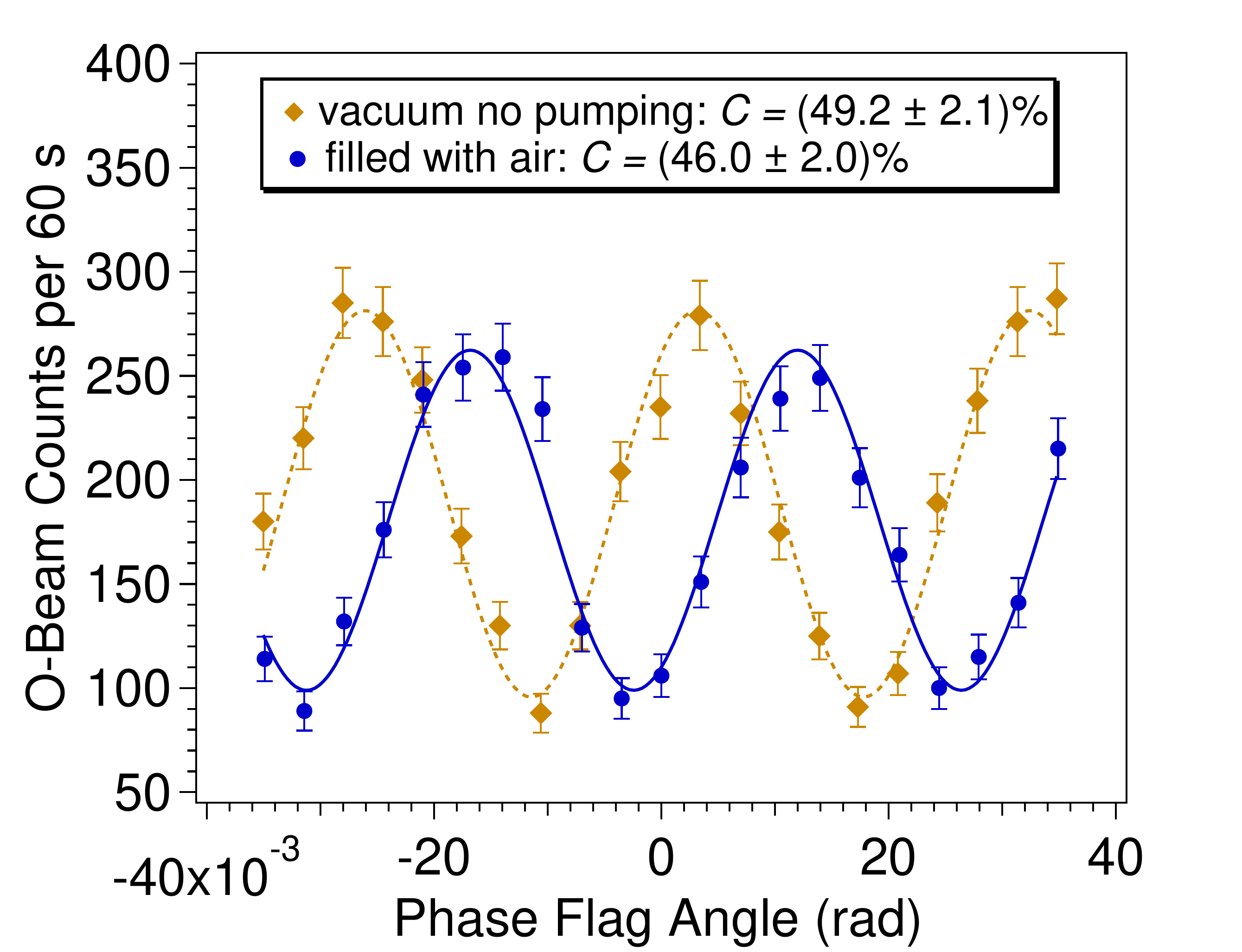}
\caption{Contrast measurements taken at ambient pressures (no vacuum or 760 Torr) (blue curve) and at $10^{-4}$ Torr with the turbo pump turned off (yellow curve). Phase uncertainties are purely statistical at the 68 \% confidence level.}
 \label{fig:contrast}
\end{figure}

We further examined the effect of the vacuum enclosure on the contrast of the NI by taking $\sim~30$~min interferogram scans. The question being investigated was whether the vacuum would induce contrast-destroying strain on the NI crystal and whether the vacuum enclosure is compatible with the rest of the NI setup. The contrast scan without the vacuum chamber was around 54~$\%$ while the contrast in an air-filled vacuum chamber was around 46~$\%$; a change in contrast most likely due to difference in the position of the interferometer before and after installation of the chamber. 
When comparing contrast with the vacuum chamber in place, contrast increases at lower pressures (Figure~\ref{fig:contrast}).

Using the turbo pump, the chamber was pumped down to its lowest achievable pressure of $4\cdot10^{-4}$ Torr. While the pump was operating, the contrast dropped to 37 \% due to mechanical vibrations caused by the pump. After the pump was turned off, with an initial pressure of $10^{-4}$ Torr the contrast increased to approximately 49~\%.

We therefore conclude that one can operate a perfect crystal neutron interferometer with high contrast and phase stability inside a vacuum chamber with isothermal walls outfitted with internal temperature control. 


%

\section{Conclusion}

Perfect crystal neutron interferometer experiments are operated in facilities that have multiple isolation mechanisms in place to reduce environmental noise. The new interferometer facility at the NCNR utilizes a compact enclosure and therefore requires smaller, more versatile methods for environmental shielding. To reduce phase instability of the interferometer due to external temperature fluctuations, we designed a vacuum chamber setup and tested using a combination of heating elements and the skew symmetric neutron interferometer. We achieved high phase stability on the level of $(0.003±0.03) \deg/hr$.  This is the first realization of a neutron interferometer experiment in vacuum, and it demonstrates that it is not only possible but highly advantageous to operate a neutron interferometer in a vacuum environment.
  

After implementing some improved vibration isolation for the vacuum pump, we will have access to a larger scope of experiments at the new beamline including the study of more exotic forces as well as long term phase stability itself and the fundamental origins of decoherence.  

\section{Acknowledgements}

We acknowledge financial support provided by NSERC \textquotedblleft Create" and \textquotedblleft Discovery" programs,  CERC and the NIST Quantum Information Program.  This work is supported by NIST and the National Science Foundation through Grant No. NSF PHY-1205342.
K. Li and W. M. Snow acknowledge the support of the Indiana University Center for Spacetime Symmetries and the Indiana University Faculty Research Support Program.
Engineering and technical support by NIST machine shop.  
We appreciate discussions with T. R. Gentile and the Neutron Beam Lifetime collaboration.

\bibliographystyle{apsrev4-1}

\bibliography{TPaper}

\end{document}